# PHYSICS OF NEUTRINO OSCILLATION

SPANDAN MONDAL

UNDERGRADUATE PROGRAMME,
INDIAN INSTITUTE OF SCIENCE, BANGALORE

Project Instructor:

## PROF. SOUROV ROY

# Contents

# 1. Introduction

In recent years, there have been several breakthroughs in neutrino physics and it has emerged as one of the most active fields of research. The Standard Model of particle physics describes neutrinos as massless, chargeless elementary particles that come in three different flavours [1]. However, recent experiments indicate that neutrinos not only have mass, but also have multiple mass eigenstates that are not identical to the flavour states, thereby indicating mixing. As an evidence of mixing, neutrinos have been observed to change from one flavour to another during their propagation – a phenomenon called neutrino oscillation.

Neutrinos can be produced from four different sources, and accordingly they are termed as solar, atmospheric, reactor and accelerator neutrinos. In the past few decades, several experiments have confirmed the event of flavour change in each of these neutrinos. Besides, the values of the parameters affecting the probabilities of neutrino oscillation have been experimentally determined in most of the cases.

In this project, we have studied the reasons and derived the probabilities of neutrino flavour change, both in vacuum and in matter. We have also studied the parameters affecting this probability. We have discussed the special case of two-neutrino oscillations. Lastly, we have discussed some basic properties of neutrinos that are reflected in the previous derivations and highlighted a few relevant open problems.

To begin with, we have also studied the relevant topics in introductory High Energy Physics and Quantum Mechanics to familiarize with the notations, units and methodologies that would be required for the subsequent project work.

# 2. Particle Physics and Quantum Mechanics: The Basics

## 2.1. Elementary Particles

During the 5th century B.C., Greek philosophers Leucippus and Democritus considered that if one keeps on cutting matter into smaller and smaller bits one will eventually end up with a fundamental bit which cannot be further subdivided. They called this smallest bit an atom. This is probably man's first attempt at finding out what the world is made of.

It goes without saying that 25 centuries worth of research has considerably refined the concept of atom, and today we have a finer understanding of what the world is made of. We know today that there is no single fundamental particle, but a number of them. Our current understanding of elementary particles can be summarized in Table 2.1 [1].

*Table 2.1: Elementary Particles*

|  | Name | Spin | Baryon Number B | Lepton Number L | Charge Q |
|---|---|---|---|---|---|
| Quarks | up (u), charm (c), top (t) | $\frac{1}{2}$ | $\frac{1}{3}$ | 0 | $+\frac{2}{3}$ |
|  | down (d), strange (s), bottom (b) | $\frac{1}{2}$ | $\frac{1}{3}$ | 0 | $-\frac{1}{3}$ |
| Leptons | electron ($e^-$), muon ($\mu^-$), tau ($\tau^-$) | $\frac{1}{2}$ | 0 | 1 | -1 |
|  | $\nu_e, \nu_\mu, \nu_\tau$ (neutrino) | $\frac{1}{2}$ | 0 | 1 | 0 |
| Guage Bosons | $\gamma$ (photon) | 1 | 0 | 0 | 0 |
|  | $W^\pm, Z$ (weak bosons) | 1 | 0 | 0 | $\pm 1, 0$ |
|  | $g_i$ (gluons) | 1 | 0 | 0 | 0 |

Spin is in the units of $\hbar$. Charges are defined such that an electron has -1 units of charge.
Each quark and lepton has a corresponding antiparticle with its B, L and Q having reversed signs.

All the particles which undergo strong interactions are called hadrons. Electrons and neutrinos do not undergo such interactions and are called leptons. While the electron is charged and can therefore undergo electromagnetic interactions, the neutrino participates exclusively in weak interactions.

Both quarks and leptons can be represented in terms of generation, wherein each generation forms a weak isospin doublet. The representation for leptons has been shown in Table 2.2.

*Table 2.2: Generations of Leptons*

| First Generation | Second Generation | Third Generation | |
|---|---|---|---|
| $\begin{pmatrix} \nu_e \\ e^- \end{pmatrix}$ | $\begin{pmatrix} \nu_\mu \\ \mu^- \end{pmatrix}$ | $\begin{pmatrix} \nu_\tau \\ \tau^- \end{pmatrix}$ | $Q = 0$ <br> $Q = -1$ |

## 2.2. Natural Units ($\hbar = c = 1$)

The two fundamental constants, viz., the Planck's constant, $h$, and the velocity of light in vacuum, $c$, appear repeatedly in expressions in relativistic quantum mechanics and high energy physics. It is convenient to use a system of units in which the use of these two constants can be avoided, thereby hugely simplifying such expressions.

This can be achieved by choosing units $\hbar = c = 1$, where $\hbar$ is the reduced Planck's constant.

$$\hbar = \frac{h}{2\pi} = 1.055 \times 10^{-34} \text{ J sec}$$
$$c = 2.998 \times 10^8 \text{ m sec}^{-1}$$

In high energy physics, quantities are measured in units of GeV, which is convenient as rest mass-energy of a proton is of the order of 1 GeV.

Keeping in mind the dimensions of $\hbar$ (ML²/T) and $c$ (L/T), it is possible to represent other quantities in terms of $\hbar$, $c$ and mass, $m$. For example, energy can be expressed as $E = mc^2$, and time can be expressed as $t = \hbar/mc^2$.

To find the conversion formulae, we might proceed as follows:

$$m \text{ kg} \equiv mc^2 \text{ Energy units}$$

$$1 \text{ kg} \equiv 1 \times (2.998 \times 10^8)^2 \text{ J}$$
$$= \frac{(2.998 \times 10^8)^2 \text{ J}}{1.6 \times 10^{-19} \frac{\text{J}}{\text{eV}}}$$
$$= 5.618 \times 10^{-35} \text{ eV}$$
$$= 5.618 \times 10^{-26} \text{ GeV}.$$

Table 2.3 shows a few other useful conversion factors.

*Table 2.3: Conversion factors for MKS to Natural units*

| Quantity | Conversion factor | Actual dimension |
|---|---|---|
| Mass | $1 \text{ kg} = 5.62 \times 10^{-26}$ GeV | $\dfrac{\text{GeV}}{c^2}$ |
| Length | $1 \text{ m} = 5.07 \times 10^{15}$ GeV$^{-1}$ | $\dfrac{\hbar c}{\text{GeV}}$ |
| Time | $1 \text{ sec} = 1.52 \times 10^{24}$ GeV$^{-1}$ | $\dfrac{\hbar}{\text{GeV}}$ |

## 2.3. Dirac's bra-ket Notation in QM

### Ket Space

In quantum mechanics, it is postulated that a physical state of a system is completely specified by a function, $\Psi(r,t)$, called the wave function. As per the notation developed by Dirac in 1939, the physical state of a quantum mechanical system can be equivalently represented by a state vector in a complex vector space [2]. Such a vector is called a ket and is denoted by $|\alpha\rangle$.

We shall state a few properties of kets:

1. $|\alpha\rangle + |\beta\rangle = |\gamma\rangle$ (1)

2. $c|\alpha\rangle = |\alpha\rangle c$ where c is any complex number. If $c = 0$, the resulting ket is a null ket. (2)

3. $A \cdot (|\alpha\rangle) = A|\alpha\rangle$ where $A$ is an operator corresponding to an observable. (3)

4. If $A|\alpha'\rangle = \alpha'|\alpha'\rangle$ then $|\alpha'\rangle$ is called an eigenket of operator $A$. (4)

### Bra Space

Now, we postulate that corresponding to every ket $|\alpha\rangle$ there exists a bra, denoted by $\langle\alpha|$ in a space called the bra space. Thus there is a one-to-one correspondence between ket and bra spaces,

$$|\alpha\rangle \overset{DC}{\leftrightarrow} \langle\alpha|$$ (5)

where DC stands for dual correspondence.

It is postulated that $c|\alpha\rangle \overset{DC}{\leftrightarrow} c^*\langle\alpha|$. (6)

We define an inner product between a bra and a ket as,
$$(\langle\beta|) \cdot (|\alpha\rangle) = \langle\beta|\alpha\rangle$$

such that,
1. $\langle\beta|\alpha\rangle = \langle\alpha|\beta\rangle^*$ (7)
2. $\langle\alpha|\alpha\rangle \geq 0$ (8)

Two kets $|\alpha\rangle$ and $|\beta\rangle$ are called orthogonal if
$$\langle\alpha|\beta\rangle = 0.$$ (9)

### Operators

An operator acts on a ket from the left and the resulting product is another ket.
$$A \cdot (|\alpha\rangle) = A|\alpha\rangle$$

An operator acts on a bra from the right and the resulting product is another bra.
$$(\langle\alpha|) \cdot A = \langle\alpha|A$$

In general, the ket $A|\alpha\rangle$ and the bra $|\alpha\rangle A$ are not dual to eachother. But,
$$A|\alpha\rangle \overset{DC}{\leftrightarrow} |\alpha\rangle A^\dagger$$ (10)

where the operator $A^\dagger$ is the Hermitian adjoint of $A$. $A$ is called Hermitian if,
$$A = A^\dagger.$$ (11)

We define the outer product of $|\beta\rangle$ and $\langle\alpha|$ as
$$(|\beta\rangle) \cdot (\langle\alpha|) = |\beta\rangle\langle\alpha|$$
which, unlike the inner operator, is not a simple number but is regarded as an operator.

## The Associative Axiom

By the associative axiom of multiplication [2],

$$(|\beta\rangle\langle\alpha|) \cdot |\gamma\rangle = |\beta\rangle \cdot (\langle\alpha|\gamma\rangle) = |\beta\rangle\langle\alpha|\gamma\rangle. \tag{12}$$

By another illustration of this axiom,

$$\langle\beta| \cdot (A|\alpha\rangle) = (\langle\beta|A) \cdot |\alpha\rangle.$$

Because both sides of this equation are same, we use a more compact expression,

$$\langle\beta|A|\alpha\rangle.$$

It can be shown that [2]

$$\langle\beta|A|\alpha\rangle = \langle\alpha|A^\dagger|\beta\rangle^* \tag{13}$$

## Matrix Representation

Hermitian operators (defined in 11) are of particular interest in quantum mechanics as they often turn out to be operators representing some physical observables. It can be proved that eigenvalues of a Hermitian operator $A$ are real and that the eigenkets of $A$ corresponding to the different eigenvalues are orthogonal [2].

Thus, if $\alpha'$ be an eigenvalue of $A$,

$$\alpha' = \alpha'^* \tag{14}$$

and,

$$\langle\alpha''|\alpha'\rangle = 0, \quad \alpha'' \neq \alpha' \tag{15}$$

Now, for an arbitrary ket $|\alpha\rangle$ in the ket space spanned by eigenkets of $A$, we can expand $|\alpha\rangle$ as,

$$|\alpha\rangle = \sum_{\alpha'} c_{\alpha'} |\alpha'\rangle \tag{16}$$

Multiplying both sides by $\langle\alpha''|$, we get

$$\langle\alpha''|\alpha\rangle = \sum_{\alpha'} c_{\alpha'} \langle\alpha''|\alpha'\rangle$$

Using (15),

$$c_{\alpha''} = \langle\alpha''|\alpha\rangle \tag{17}$$

So,

$$|\alpha\rangle = \sum_{\alpha'} |\alpha'\rangle\langle\alpha'|\alpha\rangle \tag{18}$$

Since $|\alpha\rangle$ is arbitrary, we must have,

$$\sum_{\alpha'} |\alpha'\rangle\langle\alpha'| = 1. \tag{19}$$

We might compare equation (18) to the expansion of a vector $\mathbf{V}$ in Euclidean space:

$$\mathbf{V} = \sum_i \hat{\mathbf{e}}_i (\hat{\mathbf{e}}_i \cdot \mathbf{V})$$

Clearly, $|\alpha'\rangle\langle\alpha'|$, when operates on $|\alpha\rangle$, selects that portion of ket $|\alpha\rangle$ that is parallel to $|\alpha'\rangle$. So $|\alpha'\rangle\langle\alpha'|$ is known as the projection operator along the base ket $|\alpha'\rangle$.

We can proceed similarly and using (19) twice, we can write an arbitrary operator $X$ as,

$$X = \sum_{\alpha''} \sum_{\alpha'} |\alpha''\rangle\langle\alpha''|X|\alpha'\rangle\langle\alpha'|$$

If $N$ be the dimensionality of the ket space, we have a total of $N^2$ terms. We may arrange them in an $N \times N$ matrix as,

$$X = \begin{pmatrix} \langle \alpha^{(1)}|X|\alpha^{(1)} \rangle & \langle \alpha^{(1)}|X|\alpha^{(2)} \rangle & \cdots \\ \langle \alpha^{(2)}|X|\alpha^{(1)} \rangle & \langle \alpha^{(2)}|X|\alpha^{(2)} \rangle & \cdots \\ \vdots & \vdots & \ddots \end{pmatrix} \tag{21}$$

Similarly,

$$X^\dagger = \begin{pmatrix} \langle \alpha^{(1)}|X^\dagger|\alpha^{(1)} \rangle & \langle \alpha^{(1)}|X^\dagger|\alpha^{(2)} \rangle & \cdots \\ \langle \alpha^{(2)}|X^\dagger|\alpha^{(1)} \rangle & \langle \alpha^{(2)}|X^\dagger|\alpha^{(2)} \rangle & \cdots \\ \vdots & \vdots & \ddots \end{pmatrix}$$

Using (13), it is clear that the matrix representing $X^\dagger$ is the **complex conjugate transposed** matrix of $X$.

## 2.4. Rotations in Quantum Mechanics

We recall from classical mechanics that rotations in three dimensions can be represented by real, orthogonal $3 \times 3$ matrices. The new and old components of a vector are related via the roation matrix $R$ as follows:

$$\begin{pmatrix} V_x' \\ V_y' \\ V_z' \end{pmatrix} = (R) \begin{pmatrix} V_x \\ V_y \\ V_z \end{pmatrix} \tag{22}$$

For example, a finite rotation of $\phi$ about the z-axis is achieved by using an operator $R_z(\phi)$ on a vector, where $R_z(\phi)$ is given by

$$R_z(\phi) = \begin{pmatrix} \cos\phi & -\sin\phi & 0 \\ \sin\phi & \cos\phi & 0 \\ 0 & 0 & 1 \end{pmatrix} \tag{23}$$

For an infinitesimal rotation, using Taylor's expansion and ignoring terms of order $\varepsilon^3$ and higher,

$$R_z(\varepsilon) = \begin{pmatrix} 1 - \frac{\varepsilon^2}{2} & -\varepsilon & 0 \\ \varepsilon & 1 - \frac{\varepsilon^2}{2} & 0 \\ 0 & 0 & 1 \end{pmatrix} \tag{24}$$

In quantum mechanics, we can proceed by analogy to state that there exists an operator $\mathcal{D}(R)$ in the appropriate ket space corresponding to $R$ such that,

$$|\alpha\rangle_R = \mathcal{D}(R)|\alpha\rangle \tag{25}$$

where $|\alpha\rangle$ and $|\alpha\rangle_R$ are the original and rotated systems respectively.

It can be shown that the appropriate infinitesimal operator can be written as [1]

$$U_\varepsilon = 1 - iJ\varepsilon \tag{26}$$

where $J$, the generator of rotations can be defined as a component of the angular momentum. (In classical mechanics we know that angular momentum is the generator of rotations.)

A rotation through finite angle $\theta$ may be thought to be built from $n$ successive small rotations.

$$U(\theta) = (U_\varepsilon)^n = \left(1 - \frac{iJ\theta}{n}\right)^n \xrightarrow[n\to\infty]{} e^{-iJ\theta} \tag{27}$$

## 2.5. The Klein-Gordon Equation

We recall that the Schrödinger equation for a free particle can be obtained starting from the classical, nonrelativistic energy-momentum relation,

$$E = \frac{p^2}{2m}. \qquad (28)$$

Substituting

$$E \to i\hbar \frac{\partial}{\partial t}, \quad \mathbf{p} = -i\hbar \mathbf{\nabla}, \qquad (29)$$

into (28), and shifting to natural units, we get the Schrödinger equation,

$$i \frac{\partial \psi}{\partial t} + \frac{1}{2m} \nabla^2 \psi \qquad (30)$$

where $\psi(\mathbf{x}, t)$ is a complex wavefunction.

Similarly, we start off with the relativistic energy-momentum relation (in natural units),

$$E^2 = \mathbf{p}^2 + m^2. \qquad (31)$$

Then we use the relations (29) in (31) and obtain the Klein-Gordon equation,

$$-\frac{\partial^2 \phi}{\partial t^2} + \nabla^2 \phi = m^2 \phi \qquad (32)$$

We may use the D'Alembertian operator,

$$\Box^2 \equiv \frac{\partial^2}{\partial t^2} - \nabla^2 \qquad (33)$$

in (32) to get a more compact form of the Klein-Gordon equation,

$$\left(\Box^2 + m^2\right) \phi = 0. \qquad (34)$$

Taking the free particle solution

$$\phi = N e^{i\mathbf{p}\cdot\mathbf{x} - iEt} \qquad (35)$$

and using it in (34) we get,

$$-E^2 \phi + p^2 \phi + m^2 \phi = 0 \qquad (36)$$

or, we get the energy eigenvalues of the Klein-Gordon equation as

$$E = \pm \left(\mathbf{p}^2 + m^2\right)^{1/2} \qquad (37)$$

which gives negative values of $E$ in addition to positive ones.

Dirac explained these negative solutions postulating that all negative energy states are occupied by $E < 0$ electrons. Whenever an electron is excited from $E < 0$ to $E > 0$ state, the corresponding absence of electron, or presence of a "hole" in the vacuum sea is explained as the presence of an antiparticle (positron, +e) of energy $+E$ [1]. This explains pair production of an electron and positron.

Stückelberg and Feynman suggested that a negative energy solution describes a particle that propagates backward in time or, equivalently, a positive energy antiparticle propagating forward in time. Thus an emission of a positron with energy $+E$ is the same as the absorption of an electron of energy $-E$ [1]. Mathematically, it is easy to see

$$e^{-i(-E)(-t)} = e^{-iEt}.$$

## 2.6. The Dirac Equation

As we can see from (34), the Klein-Gordon equation is quadratic in $\frac{\partial}{\partial t}$ and $\nabla$. However, Dirac wanted to write the relativistic Schrödinger equation in a form that is linear in $\frac{\partial}{\partial t}$ (implying linear in $\nabla$ as well, since the equation must be covariant). Thus, we can begin with the general form

$$H\psi = (\boldsymbol{\alpha} \cdot \mathbf{P} + \beta m)\psi \tag{38}$$

where $\alpha_i (i = 1,2,3)$ and $\beta$ are constants. However, we must satisfy (31) simultaneously. In other words,

$$H^2\psi = (\mathbf{P}^2 + m^2)\psi \tag{39}$$

and (38) must together represent the Dirac Equation.

Starting from (38), we write (in Einstein summation convention)

$$H^2\psi = (\alpha_i P_i + \beta m)(\alpha_j P_j + \beta m)\psi$$
$$= \left(\alpha_i^2 P_i^2 + (\alpha_i\alpha_j + \alpha_j\alpha_i)P_i P_j + (\alpha_i\beta + \beta\alpha_i)P_i m + \beta^2 m^2\right)\psi$$

Comparing with (39),

- $\alpha_i^2 = \beta^2 = 0$ (40)
- $\alpha_i\alpha_j + \alpha_j\alpha_i = \alpha_i\beta + \beta\alpha_i = 0$. Hence $\alpha_i$'s and $\beta$ anticommute with one another. (41)

By (41) $\alpha_i$'s and $\beta$ cannot be numbers and must be represented by matrices.

The lowest dimensionality matrices satisfying these are $4 \times 4$. They are not unique. One representation, known as Dirac-Pauli representation is as follows [1]

$$\boldsymbol{\alpha} = \begin{pmatrix} 0 & \boldsymbol{\sigma} \\ \boldsymbol{\sigma} & 0 \end{pmatrix}, \quad \beta = \begin{pmatrix} I & 0 \\ 0 & -I \end{pmatrix}$$

where $\boldsymbol{\sigma}$ are the Pauli matrices [1]:

$$\sigma_1 = \begin{pmatrix} 0 & 1 \\ 1 & 0 \end{pmatrix}, \quad \sigma_2 = \begin{pmatrix} 0 & -i \\ i & 0 \end{pmatrix}, \quad \sigma_3 = \begin{pmatrix} 1 & 0 \\ 0 & -1 \end{pmatrix}$$

Making necessary operator substitutions given by (29), we can rewrite (38) as

$$i\frac{\partial\psi}{\partial t} = -i\boldsymbol{\alpha} \cdot \nabla\psi + m\psi \tag{42}$$

with $\alpha_i$'s and $\beta$ satisfying (40) and (41).

Multiplying (42) from the left by $\beta$, we get

$$i\beta\frac{\partial\psi}{\partial t} = -i\beta\boldsymbol{\alpha} \cdot \nabla\psi + \beta m\psi$$

which can be rewritten as

$$(i\gamma^\mu\partial_\mu - m)\psi = 0 \tag{43}$$

where $\gamma^\mu \equiv (\beta, \beta\boldsymbol{\alpha})$ and $\partial_\mu = \left(\frac{\partial}{\partial t}, \nabla\right)$ (in four vector notation). $\gamma^\mu$'s are known as the Dirac $\gamma$ matrices.

Since $\gamma^0 = \beta$,

$$\gamma^{0\dagger} = \gamma^0 \tag{44}$$
$$(\gamma^0)^2 = I \tag{45}$$

But, for $\gamma^k$, $k = 1,2,3$,

$$\gamma^{k\dagger} = (\beta\alpha^k)^\dagger = \alpha^k\beta = -\gamma^k \tag{46}$$

We take the Hermitian conjugate of the Dirac equation

$$i\gamma^0\frac{\partial\psi}{\partial t} + i\gamma^k\frac{\partial\psi}{\partial x^k} - m\psi = 0$$

where $k = 1,2,3$, and using (44) and (46) we get,

$$-i\frac{\partial\psi^\dagger}{\partial t}\gamma^0 - i\frac{\partial\psi^\dagger}{\partial x^k}(-\gamma^k) - m\psi^\dagger = 0 \tag{47}$$

Multiplying (47) by $\gamma^0$ from the right,

$$-i\frac{\partial\psi^\dagger}{\partial t}\gamma^0\gamma^0 + i\frac{\partial\psi^\dagger}{\partial x^k}\gamma^k\gamma^0 - m\psi^\dagger\gamma^0 = 0$$

Or,

$$-i\frac{\partial\psi^\dagger}{\partial t}\gamma^0\gamma^0 - i\frac{\partial\psi^\dagger}{\partial x^k}\gamma^0\gamma^k - m\psi^\dagger\gamma^0 = 0$$

Or,

where $\bar{\psi} = \psi^\dagger \gamma^0$. Or,

$$-i\frac{\partial\bar{\psi}}{\partial t}\gamma^0 - i\frac{\partial\bar{\psi}}{\partial x^k}\gamma^k - m\bar{\psi} = 0$$

$$i\partial_\mu \bar{\psi}\gamma^\mu + m\bar{\psi} = 0 \tag{48}$$

Multiplying (43) by $\bar{\psi}$ from the left and (48) by $\psi$ from the right and adding,

$$\bar{\psi}\gamma^\mu \partial_\mu \psi + (\partial_\mu \bar{\psi})\gamma^\mu \psi = \partial_\mu(\bar{\psi}\gamma^\mu \psi) = 0 \tag{49}$$

which gives the continuity equation.

Thus the four-vector current density of, say, an electron of charge –e can be given by [1]

$$J^\mu = -e\bar{\psi}\gamma^\mu \psi \tag{50}$$

## 2.7. Bilinear covariants and Weak Interactions

We can further generalize current densities by using products of $\gamma$-matrices instead of just one as in (50). We can list a number of bilinear quantities of the form

$$(\bar{\psi})(4 \times 4)(\psi).$$

For notational simplicity, we define

$$\gamma^5 \equiv i\gamma^0 \gamma^1 \gamma^2 \gamma^3. \tag{51}$$

*Table 2.4: Bilinear Covariants*

| Scalar | $\bar{\psi}\psi$ |
|---|---|
| Vector | $\bar{\psi}\gamma^\mu \psi$ |
| Tensor | $\bar{\psi}\sigma^{\mu\nu}\psi$ |
| Axial Vector | $\bar{\psi}\gamma^5 \gamma^\mu \psi$ |
| Pseudoscalar | $\bar{\psi}\gamma^5 \psi$ |

Experiments indicate that leptons participate in weak interactions in a special combination of two of the bilinear covariants. For electron and an electron-flavoured neutrino [1],

$$J^\mu = -\frac{1}{2}\bar{\psi}_e \gamma^\mu (1 - \gamma^5)\psi_\nu$$

which takes a form of $V - A$ (Vector 'minus' Axial Vector, $(\gamma^\mu - \gamma^\mu \gamma^5)$).

# 3. Neutrino Flavours, Mass Eigenstates and Mixing

## 3.1. Introduction

As discussed earlier, neutrinos come in three flavours, electron neutrinos ($\nu_e$), muon neutrinos ($\nu_\mu$), and tau neutrinos ($\nu_\tau$). Each flavour is, of course, associated with a corresponding antiparticle called an antineutrino. The Standard Model states that neutrinos are massless and chargeless, and only undergo weak interactions.

However, it has been observed that neutrinos can change their flavours during their travel. That is, a neutrino which was generated with a certain flavour might end up having a different flavour after travelling some distance. Such flavour changes require, as we shall show in subsequent sections, that neutrino flavours have different masses with significant mixing. This, in turn, implies that neutrinos are not massless and, therefore, can participate in gravitational forces as well.

## 3.2. Leptonic Mixing

To begin with, we shall assume that neutrinos have masses. Thus there is a spectrum of neutrino mass eigenstates, $\nu_i, i = 1, 2, ...$, each with mass $m_i$. We can, however, distinguish between neutrinos in terms of their flavours as well. We define three flavour states, $\nu_\alpha, \alpha = e, \mu, \tau$, as observed experimentally.

Mixing may be described with the observation that each of the three flavours of neutrinos can be expressed as a superposition of mass eigenstates. To understand this experimentally, consider the leptonic decay

$$W^+ \to \nu_i + \overline{\ell_\alpha} \qquad (52)$$

where $\ell_\alpha$ is a charged lepton of flavour $\alpha$. Mixing implies that every time the above decay produces a particular $\overline{\ell_\alpha}$, the accompanying neutrino mass eigenstate is not the same $\nu_i$, but can be any $\nu_i$ even if the lepton has a fixed flavour. Thus, each $\nu_\alpha$ is actually a superposition of several eigenstates $\nu_i$'s, of which, of course, only one state can be discerned during a single observation.

Thus, we can write a flavour state $|\nu_\alpha\rangle$ in the form [3]:

$$|\nu_\alpha\rangle = \sum_i U_{\alpha i} |\nu_i\rangle \qquad (53)$$

The $U_{\alpha i}$'s may be written in a matrix form, called leptonic mixing matrix. A typical leptonic mixing matrix assuming $i = 1, 2, 3$ and $\alpha = e, \mu, \tau$ would look like

$$U = \begin{pmatrix} U_{e1} & U_{e2} & U_{e3} \\ U_{\mu 1} & U_{\mu 2} & U_{\mu 3} \\ U_{\tau 1} & U_{\tau 2} & U_{\tau 3} \end{pmatrix}$$

The Standard Model states $U$ is unitary. Thus,

$$UU^\dagger = U^\dagger U = I \qquad (54)$$

Inverting (53), we can express each mass eigenstate as a superposition of flavours

$$|\nu_i\rangle = \sum_\alpha U^*_{\alpha i} |\nu_\alpha\rangle \qquad (55)$$

The corresponding mixing matrix is, of course, $U^\dagger$.

Assuming there are 3 mass eigenstates, the Lagrangian can be expressed in mass eigenstate (basis) terms as:

$$\mathcal{L} = \bar{\nu} m \nu$$
$$= (\bar{\nu}_1 \quad \bar{\nu}_2 \quad \bar{\nu}_3) \underbrace{\begin{pmatrix} m_1 & 0 & 0 \\ 0 & m_2 & 0 \\ 0 & 0 & m_3 \end{pmatrix}}_{M_D} \begin{pmatrix} \nu_1 \\ \nu_2 \\ \nu_3 \end{pmatrix} \qquad (56)$$

$\mathcal{L}$ can also be expressed in flavour basis, as

$$\mathcal{L} = (\bar{\nu}_\alpha \quad \bar{\nu}_\mu \quad \bar{\nu}_\tau) \underbrace{\begin{pmatrix} M_{11} & M_{12} & M_{13} \\ M_{21} & M_{22} & M_{23} \\ M_{31} & M_{32} & M_{33} \end{pmatrix}}_{M} \begin{pmatrix} \nu_\alpha \\ \nu_\mu \\ \nu_\tau \end{pmatrix} \qquad (57)$$

Clearly, $M$ would be diagonal if there was no mixing. Using (53) and (55) in (57), and assuming $M$ to be symmetric, it follows

$$M_D = U^\dagger M U \qquad (58)$$

From (53) and (55) it is clear that the flavour-$\alpha$ fraction in $\nu_i$, and, equivalently, mass-$i$ fraction in $\nu_\alpha$ is $|U_{\alpha i}|^2$. This is, therefore, the probability that the neutrino will have mass $m_i$ when a $\overline{\ell_\alpha}$ is observed in the decay (52).

# 4. Neutrino Oscillation in Vacuum

## 4.1. Detecting a flavour change

The term 'neutrino oscillation' refers to the phenomenon of neutrino flavour change, for reasons stated later on. However, as discussed earlier, neutrinos participate in weak interactions only (and very weakly in gravity, owing to their extremely small masses), which makes it difficult to detect them. But the charged lepton produced alongside a neutrino can be easily detected and its flavour can be identified. Thus we can determine the flavour of the produced neutrino, say $\alpha$.

Similarly, at the end of a path of length $L$, the neutrino reacts with the detector and produces a charged lepton. Determining the flavour of the charged lepton at the detector, we can find the final flavour of the neutrino, say $\beta$.

If $\alpha \neq \beta$, then the neutrino has changed its flavour in its journey. This neutrino flavour change, $\nu_\alpha \to \nu_\beta$ is a quantum mechanical effect, and we can try and find out the probability of such a change, $P(\nu_\alpha \to \nu_\beta)$.

## 4.2. Finding the oscillation probability

Since each $\nu_\alpha$ is a superposition of $\nu_i$'s, we have to individually add the contribution from each travelling $\nu_i$ while finding $P(\nu_\alpha \to \nu_\beta)$.

We shall first find the amplitude, denoted by $\text{Amp}(\nu_\alpha \to \nu_\beta)$. Each $\nu_i$'s contribution will depend on three factors [3]:

- The amplitude for $\nu_i$ when a $\overline{\ell_\alpha}$ is produced at the source. As explained earlier, this is given by $U_{\alpha i}$.
- The amplitude for $\nu_i$ to propagate from source to detector. Let this be $\text{Prop}(\nu_i)$.
- The amplitude for $\nu_i$ when a $\overline{\ell_\beta}$ is detected at the detector. This is given by $U_{\beta i}^*$.

Thus, the amplitude of flavour change $\nu_\alpha \to \nu_\beta$,

$$\text{Amp}(\nu_\alpha \to \nu_\beta) = \sum_i U_{\alpha i} \text{Prop}(\nu_i) U_{\beta i}^*$$

*(59)*

We shall now find the value of $\text{Prop}(\nu_i)$. In the rest frame of the neutrino, its state vector as a function of time $\tau$, follows the simple Schrödinger equation,

$$i\frac{\partial}{\partial \tau}|\nu_i(\tau)\rangle = m_i|\nu_i(\tau)\rangle$$

whose solution is given by

$$|\nu_i(\tau)\rangle = e^{-m_i \tau}|\nu_i(0)\rangle.$$

The amplitude of $\nu_i$ travelling for time $\tau_0$ is given by

$$\langle \nu_i(0)|\nu_i(\tau_0)\rangle = e^{-m_i \tau_0}$$

Thus if $\tau_i$ be the time taken by the $\nu_i$ neutrino to travel from its source to the detector in its rest frame (proper time $\tau_i$), then,

$$\text{Prop}_{Rest}(\nu_i) = \langle \nu_i(0)|\nu_i(\tau_i)\rangle = e^{-m_i \tau_i} \qquad (60)$$

However, we need Prop($\nu_i$) in the lab frame. So we shall need to use a Lorentz transform to find the corresponding expression in the lab frame. The lab frame variables are [3]:
- Distance between source and detector, $L$.
- Laboratory-frame time, $t$.
- Energy of mass eigenstate $\nu_i$, $E_i$.
- Momentum of mass eigenstate $\nu_i$, $p_i$.

By Lorentz invariance,
$$m_i \tau_i = E_i t - p_i L \qquad (61)$$

To eliminate $p_i$, we use the relation
$$p_i = \sqrt{E^2 - m_i^2} \cong E - \frac{m_i^2}{2E} \qquad (62)$$

where we have used the fact that rest energy of neutrinos are extremely tiny, $m_i^2 \ll E^2$. Using (62) in (61),
$$\begin{aligned} m_i \tau_i &\cong Et - EL + \frac{m_i^2}{2E} L \\ &= E(t-L) + \frac{m_i^2}{2E} L. \end{aligned} \qquad (63)$$

Here we have used the same $E$ for different mass eigenstates. It can be justified as follows. Suppose two different components, $\nu_i$ and $\nu_j$, have different energies $E_i$ and $E_j$. By the time they reach the detector, they have phases of $e^{-iE_i t}$ and $e^{-iE_j t}$ respectively, where $t$ is travel time in lab frame. Thus the detector detects an interference caused by two components with a phase difference of $e^{-i(E_i - E_j)t}$, which vanishes for an average over time for $i \neq j$ [3]. Thus only components with same energy are detected.

The term $E(t-L)$ is, therefore, common to every interfering mass eigenstate. Thus, considering only the $i$-dependent part
$$\text{Prop}(\nu_i) = e^{-i\frac{m_i^2}{2E}L}. \qquad (64)$$

Thus, (59) can be finally written as
$$\text{Amp}(\nu_\alpha \to \nu_\beta) = \sum_i U_{\alpha i} e^{-i\frac{m_i^2}{2E}L} U_{\beta i}^*. \qquad (65)$$

Next, we shall find $P(\nu_\alpha \to \nu_\beta)$ from (65) [3].
$$\begin{aligned} P(\nu_\alpha \to \nu_\beta) &= |\text{Amp}(\nu_\alpha \to \nu_\beta)|^2 \\ &= \left(\sum_i U_{\alpha i} e^{-i\frac{m_i^2}{2E}L} U_{\beta i}^*\right)^* \left(\sum_j U_{\alpha j} e^{-i\frac{m_j^2}{2E}L} U_{\beta j}^*\right) \\ &= \sum_i \sum_j U_{\alpha i}^* U_{\beta i} U_{\alpha j} U_{\beta j}^* e^{i\frac{L}{2E}(m_j^2 - m_i^2)} \\ &= \sum_i U_{\alpha i}^* U_{\beta i} U_{\alpha i} U_{\beta i}^* + \sum_{i \neq j} U_{\alpha i}^* U_{\beta i} U_{\alpha j} U_{\beta j}^* e^{i\frac{L}{2E}\Delta m_{ji}^2} \end{aligned} \qquad (66)$$

where
$$\Delta m_{ji}^2 = (m_j^2 - m_i^2) \qquad (67)$$

We derive the following identity:
$$e^{iA} = \cos A + i \sin A$$

$$= 1 - 2\sin^2\frac{A}{2} + i\sin A$$

(68)

We shall write (66) as a sum of four sums by expanding $e^{i\frac{L}{2E}\Delta m_{ji}^2}$ using (68), as

$$P(\nu_\alpha \to \nu_\beta) = \overbrace{\sum_i U_{\alpha i}^* U_{\beta i} U_{\alpha i} U_{\beta i}^*}^{P_1} + \overbrace{\sum_{i\neq j} U_{\alpha i}^* U_{\beta i} U_{\alpha j} U_{\beta j}^*}^{P_2} - \overbrace{2\sum_{i\neq j} U_{\alpha i}^* U_{\beta i} U_{\alpha j} U_{\beta j}^* \sin^2\left(\Delta m_{ji}^2 \frac{L}{4E}\right)}^{P_3}$$
$$+ \underbrace{i\sum_{i\neq j} U_{\alpha i}^* U_{\beta i} U_{\alpha j} U_{\beta j}^* \sin\left(\Delta m_{ji}^2 \frac{L}{2E}\right)}_{P_4}.$$

(69)

Next, we shall evaluate each part of the expression individually.

$$P_3 = \sum_{i\neq j} U_{\alpha i}^* U_{\beta i} U_{\alpha j} U_{\beta j}^* \sin^2\left(\Delta m_{ji}^2 \frac{L}{4E}\right)$$
$$= \sum_{i>j} U_{\alpha i}^* U_{\beta i} U_{\alpha j} U_{\beta j}^* \sin^2\left(\Delta m_{ij}^2 \frac{L}{4E}\right) + \sum_{i<j} U_{\alpha i}^* U_{\beta i} U_{\alpha j} U_{\beta j}^* \sin^2\left(\Delta m_{ji}^2 \frac{L}{4E}\right)$$
$$= \sum_{i>j} U_{\alpha i}^* U_{\beta i} U_{\alpha j} U_{\beta j}^* \sin^2\left(\Delta m_{ij}^2 \frac{L}{4E}\right) + \sum_{i>j} U_{\alpha j}^* U_{\beta j} U_{\alpha i} U_{\beta i}^* \sin^2\left(\Delta m_{ij}^2 \frac{L}{4E}\right)$$
$$= \sum_{i>j} \sin^2\left(\Delta m_{ij}^2 \frac{L}{4E}\right) \left(U_{\alpha i}^* U_{\beta i} U_{\alpha j} U_{\beta j}^* + U_{\alpha j}^* U_{\beta j} U_{\alpha i} U_{\beta i}^*\right)$$
$$= \sum_{i>j} \sin^2\left(\Delta m_{ij}^2 \frac{L}{4E}\right) \left(U_{\alpha i}^* U_{\beta i} U_{\alpha j} U_{\beta j}^* + U_{\alpha i} U_{\beta i}^* U_{\alpha j}^* U_{\beta j}\right)$$
$$= \sum_{i>j} \sin^2\left(\Delta m_{ij}^2 \frac{L}{4E}\right) \left[U_{\alpha i}^* U_{\beta i} U_{\alpha j} U_{\beta j}^* + (U_{\alpha i}^* U_{\beta i} U_{\alpha j} U_{\beta j}^*)^*\right]$$
$$= 2\sum_{i>j} \Re(U_{\alpha i}^* U_{\beta i} U_{\alpha j} U_{\beta j}^*) \sin^2\left(\Delta m_{ij}^2 \frac{L}{4E}\right)$$

(70)

where $\Re(Z)$ denotes the real part of a complex number $Z$.

Note, in the second step we have replaced the first $\Delta m_{ji}^2$ with $\Delta m_{ij}^2$ keeping the sign same as $\sin^2$ is an even function.

We shall proceed similarly for $P_4$:

$$P_4 = \sum_{i\neq j} U_{\alpha i}^* U_{\beta i} U_{\alpha j} U_{\beta j}^* \sin\left(\Delta m_{ji}^2 \frac{L}{2E}\right)$$
$$= -\sum_{i>j} U_{\alpha i}^* U_{\beta i} U_{\alpha j} U_{\beta j}^* \sin\left(\Delta m_{ij}^2 \frac{L}{2E}\right) + \sum_{i<j} U_{\alpha i}^* U_{\beta i} U_{\alpha j} U_{\beta j}^* \sin\left(\Delta m_{ji}^2 \frac{L}{2E}\right)$$
$$= -\sum_{i>j} U_{\alpha i}^* U_{\beta i} U_{\alpha j} U_{\beta j}^* \sin\left(\Delta m_{ij}^2 \frac{L}{2E}\right) + \sum_{i>j} U_{\alpha j}^* U_{\beta j} U_{\alpha i} U_{\beta i}^* \sin\left(\Delta m_{ij}^2 \frac{L}{2E}\right)$$
$$= \sum_{i>j} \sin\left(\Delta m_{ij}^2 \frac{L}{2E}\right) \left(U_{\alpha j}^* U_{\beta j} U_{\alpha i} U_{\beta i}^* - U_{\alpha i}^* U_{\beta i} U_{\alpha j} U_{\beta j}^*\right)$$
$$= \sum_{i>j} \sin\left(\Delta m_{ij}^2 \frac{L}{2E}\right) \left[(U_{\alpha i}^* U_{\beta i} U_{\alpha j} U_{\beta j}^*)^* - U_{\alpha i}^* U_{\beta i} U_{\alpha j} U_{\beta j}^*\right]$$
$$= \sum_{i>j} \sin\left(\Delta m_{ij}^2 \frac{L}{2E}\right) \left[-2i\Im(U_{\alpha i}^* U_{\beta i} U_{\alpha j} U_{\beta j}^*)\right]$$

$$= -2i \sum_{i>j} \Im(U^*_{\alpha i} U_{\beta i} U_{\alpha j} U^*_{\beta j}) \sin\left(\Delta m^2_{ij} \frac{L}{2E}\right).$$

(71)

where $\Im(Z)$ denotes the imaginary part of a complex number $Z$. In the second step we have replaced the first $\Delta m^2_{ji}$ with $\Delta m^2_{ij}$ and changed the sign of that term as sin is an odd function.

We shall evaluate the two terms $P_1$ and $P_2$ jointly.

$$P_1 + P_2 = \sum_i U^*_{\alpha i} U_{\beta i} U_{\alpha i} U^*_{\beta i} + \sum_{i \neq j} U^*_{\alpha i} U_{\beta i} U_{\alpha j} U^*_{\beta j}$$

$$= \sum_i \sum_j U^*_{\alpha i} U_{\beta i} U_{\alpha j} U^*_{\beta j}$$

$$= \sum_i (U^*_{\alpha i} U_{\beta i}) \sum_j (U_{\alpha j} U^*_{\beta j})$$

$$= \left| \sum_i U_{\alpha i} U^*_{\beta i} \right|^2.$$

(72)

To evaluate this, we shall use the unitary property of $U$ (54). Assuming three mass eigenstates, $U$ and $U^\dagger$ can be written as

$$U = \begin{pmatrix} U_{e1} & U_{e2} & U_{e3} \\ U_{\mu 1} & U_{\mu 2} & U_{\mu 3} \\ U_{\tau 1} & U_{\tau 2} & U_{\tau 3} \end{pmatrix}, \quad U^\dagger = \begin{pmatrix} U^*_{e1} & U^*_{\mu 1} & U^*_{\tau 1} \\ U^*_{e2} & U^*_{\mu 2} & U^*_{\tau 2} \\ U^*_{e3} & U^*_{\mu 3} & U^*_{\tau 3} \end{pmatrix}$$

(73)

$$\therefore UU^\dagger = \begin{pmatrix} \sum_i U_{ei} U^*_{ei} & \sum_i U_{ei} U^*_{\mu i} & \sum_i U_{ei} U^*_{\tau i} \\ \sum_i U_{\mu i} U^*_{ei} & \sum_i U_{\mu i} U^*_{\mu i} & \sum_i U_{\mu i} U^*_{\tau i} \\ \sum_i U_{\tau i} U^*_{ei} & \sum_i U_{\tau i} U^*_{\mu i} & \sum_i U_{\tau i} U^*_{\tau i} \end{pmatrix} = \begin{pmatrix} 1 & 0 & 0 \\ 0 & 1 & 0 \\ 0 & 0 & 1 \end{pmatrix}$$

(74)

Or, in other words,

$$\sum_i U_{\alpha i} U^*_{\beta i} = \begin{cases} 1 & \text{if } \alpha = \beta \\ 0 & \text{if } \alpha \neq \beta \end{cases}$$
$$= \delta_{\alpha\beta}$$

(75)

where $\delta$ is the Kronecker delta function.

Thus, using the result (75) in (72), we get

$$P_1 + P_2 = \delta_{\alpha\beta}.$$

(76)

Now, putting (71), (72) and (76) in (69), we get

$$P(\nu_\alpha \to \nu_\beta) = \delta_{\alpha\beta} - 2 \cdot 2 \sum_{i>j} \Re(U^*_{\alpha i} U_{\beta i} U_{\alpha j} U^*_{\beta j}) \sin^2\left(\Delta m^2_{ij} \frac{L}{4E}\right)$$

$$+ i \left[ -2i \sum_{i>j} \Im(U^*_{\alpha i} U_{\beta i} U_{\alpha j} U^*_{\beta j}) \sin\left(\Delta m^2_{ij} \frac{L}{2E}\right) \right]$$

$$= \delta_{\alpha\beta} - 4 \sum_{i>j} \Re(U^*_{\alpha i} U_{\beta i} U_{\alpha j} U^*_{\beta j}) \sin^2\left(\Delta m^2_{ij} \frac{L}{4E}\right) + 2 \sum_{i>j} \Im(U^*_{\alpha i} U_{\beta i} U_{\alpha j} U^*_{\beta j}) \sin\left(\Delta m^2_{ij} \frac{L}{2E}\right)$$

(77)

## 4.3. Analysis and Discussions

1. To extend (77) to antineutrinos, we assume that CPT invariance holds true. Under this assumption, the process $\overline{\nu_\alpha} \to \overline{\nu_\beta}$ is the CPT-mirror image of $\nu_\beta \to \nu_\alpha$
$$P(\overline{\nu_\alpha} \to \overline{\nu_\beta}) = P(\nu_\beta \to \nu_\alpha).$$

Interchanging $\alpha$ and $\beta$ in $U^*_{\alpha i}U_{\beta i}U_{\alpha j}U^*_{\beta j}$ gives $U^*_{\beta i}U_{\alpha i}U_{\beta j}U^*_{\alpha j}$, which is nothing but the complex conjugate of $U^*_{\alpha i}U_{\beta i}U_{\alpha j}U^*_{\beta j}$. Thus in the expression of $P(\nu_\beta \to \nu_\alpha)$, we simply need to reverse the sign of the term containing $\Im(U^*_{\alpha i}U_{\beta i}U_{\alpha j}U^*_{\beta j})$ in (77).

Thus,
$$P(\overline{\nu_\alpha} \to \overline{\nu_\beta}) = P(\nu_\beta \to \nu_\alpha)$$
$$= \delta_{\alpha\beta} - 4\sum_{i>j} \Re(U^*_{\alpha i}U_{\beta i}U_{\alpha j}U^*_{\beta j}) \sin^2\left(\Delta m^2_{ij}\frac{L}{4E}\right) - 2\sum_{i>j} \Im(U^*_{\alpha i}U_{\beta i}U_{\alpha j}U^*_{\beta j})\sin\left(\Delta m^2_{ij}\frac{L}{2E}\right)$$
(78)

2. As the probability of neutrino flavour change is a sum of sinusoidal and sine-squared functions, it necessarily oscillates with the value of $\frac{L}{E}$. Hence the term "Neutrino Oscillation" is used.

3. If all neutrinos were massless, then the mass squared difference, $\Delta m^2_{ij} = 0$. Which reduces (77) and (78) to
$$P(\overline{\nu_\alpha} \to \overline{\nu_\beta}) = P(\nu_\alpha \to \nu_\beta) = \delta_{\alpha\beta}$$
which gives a zero probability of the event $\nu_\alpha \to \nu_\beta$ for $\alpha \neq \beta$.
This means the fact that neutrinos have been observed to undergo flavour changes in vacuum implies that neutrinos are not massless and, additionally, mass eigenstates are not degenerate.

4. Since the entire calculation has been done for the case of neutrinos travelling in vacuum, it is clear that the phenomenon of flavour change does not arise from interactions with matter, but arises from the time evolution of a neutrino itself.

5. If there was no leptonic mixing, all off-diagonal terms in $U_{\alpha i}$ would be zeroes. Thus at least one among $U^*_{\alpha i}$ and $U_{\alpha j}$ in $U^*_{\alpha i}U_{\beta i}U_{\alpha j}U^*_{\beta j}$ is a zero for $i > j$. Which again reduces (77) and (78) to
$$P(\overline{\nu_\alpha} \to \overline{\nu_\beta}) = P(\nu_\alpha \to \nu_\beta) = \delta_{\alpha\beta}$$
This means the fact that neutrinos have been observed to undergo flavour changes in vacuum implies leptonic mixing.

6. We shall include the $\hbar$ and $c$ terms in $\Delta m^2_{ij}\frac{L}{4E}$ to get a quantitative estimate. This can be done by dimensional analysis. Since $\Delta m^2_{ij}\frac{L}{4E}$ is an argument for a trigonometric function, it must be dimensionless. $\Delta m^2_{ij}\frac{L}{4E}$ has a dimension of $\left[\frac{M^2L}{\frac{ML^2}{s^2}}\right] = \left[\frac{Ms^2}{L}\right]$. From observation, $\Delta m^2_{ij}\frac{L}{4E}\frac{c^3}{\hbar}$ has a dimension $\left[\frac{Ms^2}{L}\right]\left[\frac{L^3s}{s^3ML^2}\right] = [1]$. So, $\Delta m^2_{ij}\frac{L}{4E}\frac{c^3}{\hbar}$ is the correct expression in S.I.

To be able to write the numerical values of mass in eV, length in km and energy in GeV, we convert to expression from SI, to suitable units using conversion factors in Table 1.3:
$$\Delta m^2_{ij}\frac{L}{4E}\bigg|_{Natural} = \Delta m^2_{kg}\frac{L_{meter}}{4E_{Joule}}\frac{c^3}{\hbar}$$
$$= \Delta m^2_{eV}\frac{10^3}{(5.61 \times 10^{35})^2} \times \frac{L_{km}}{4E_{GeV}}\frac{1}{1.6 \times 10^{-10}} \times \frac{(2.998 \times 10^8)^3}{1.055 \times 10^{-34}}$$

$$= 1.27 \Delta m_{eV}^2 \frac{L_{km}}{E_{GeV}}$$

$$= 1.27 \Delta m_{ij}^2 (\text{eV}^2) \frac{L(\text{km})}{E(\text{GeV})}.$$

(79)

where, for example, $\Delta m_{ij}^2(\text{eV}^2)$ and $\Delta m_{eV}^2$ are both used to mean $\Delta m_{ij}^2$ expressed in electron volts. Experimentally, the sinusoidal terms are discernable as long as their arguments are of the order of unity or larger. Thus, if we need a $\Delta m_{ij}^2$ sensitivity of, say, $10^{-5}$ with $E = 1$ GeV, we shall need a path of $L = 10^5$ km.

7. Equation (77) and (78) contain the term $\Delta m_{ij}^2$, but do not contain the mass of each mass eigenstate explicitly. Hence, although we can find out the squared-mass splitting from neutrino oscillation experiments, we cannot find out the mass of each eigenstate.

8. Suppose only two mass eigenstates $\nu_1$ and $\nu_2$ were significant, with $\Delta m_{21}^2 = \Delta m^2$. Correspondingly, the two flavour states are $\nu_e$ and $\nu_\mu$. We expect a corresponding $2 \times 2$ mixing matrix $U$ to exist, which must be unitary. A $2 \times 2$ unitary matrix has 1 rotation angle and 3 phase factors. It can be shown, that phase factors have no effect on oscillation and hence can be omitted [3]. Thus, the only possible unitary matrix with one angle parameter is:

$$U = \begin{pmatrix} U_{e1} & U_{e2} \\ U_{\mu 1} & U_{\mu 2} \end{pmatrix} = \begin{pmatrix} \cos\theta & \sin\theta \\ -\sin\theta & \cos\theta \end{pmatrix}$$

(80)

and consequently,

$$U^\dagger = \begin{pmatrix} U_{e1}^* & U_{\mu 1}^* \\ U_{e2}^* & U_{\mu 2}^* \end{pmatrix} = \begin{pmatrix} \cos\theta & -\sin\theta \\ \sin\theta & \cos\theta \end{pmatrix}$$

Using this in (77),

$$4 U_{\alpha 2}^* U_{\beta 2} U_{\alpha 1} U_{\beta 1}^* = -4 \sin\theta \cos\theta \cos\theta \sin\theta$$
$$= -\sin^2 2\theta.$$

Thus, from (77), for $\alpha \neq \beta$,

$$P(\nu_\alpha \to \nu_\beta) = \delta_{\alpha\beta} - (-\sin^2 2\theta)\sin^2\left(\Delta m_{ij}^2 \frac{L}{4E}\right) + 2 \times 0 \times \sin\left(\Delta m_{ij}^2 \frac{L}{2E}\right)$$
$$= \sin^2 2\theta \sin^2\left(\Delta m_{ij}^2 \frac{L}{4E}\right)$$

(81)

Note, the same equation applies for $P(\overline{\nu_\alpha} \to \overline{\nu_\beta})$ since the last term vanishes for a real $U$ matrix.

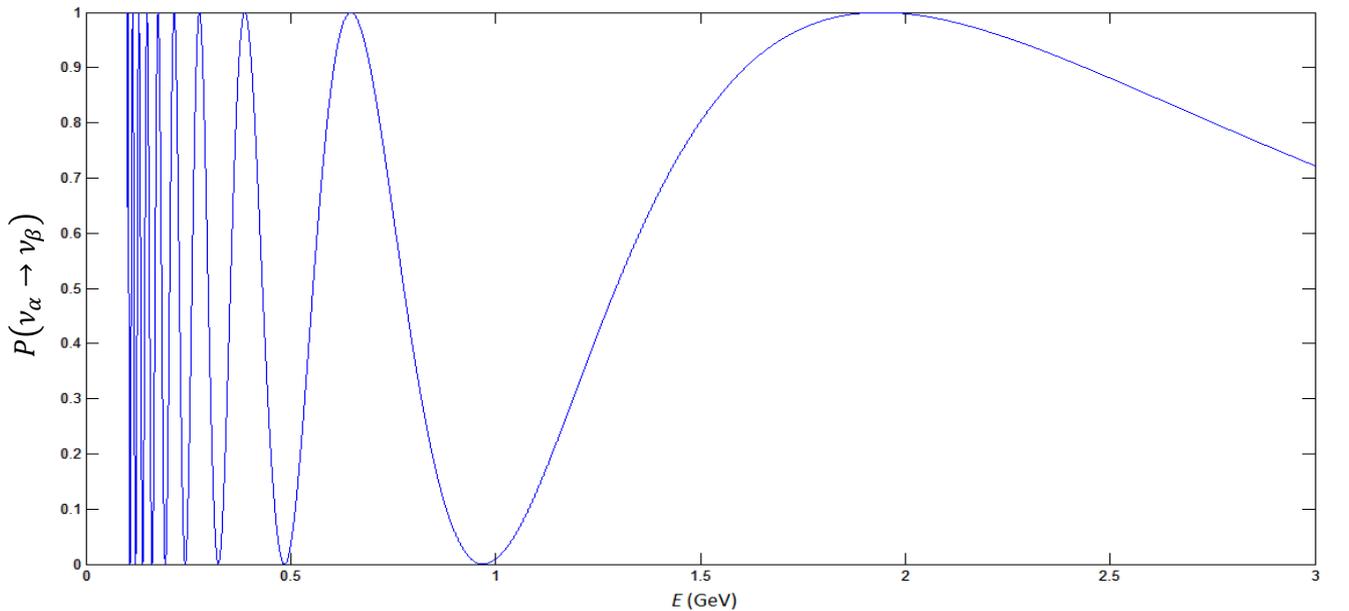

Figure 4.1: Probability vs. Energy plot with L = 1000 km, $\Delta m_{ij}^2 = 2.4 \times 10^{-3} eV^2$ at $\theta = 45°$.

From (81) we plot $P(\nu_\alpha \to \nu_\beta)$ vs. Energy $E$ for the two-neutrino case. For a definite $L$, the probability will vary with $E$ somewhat as in Fig. 4.1. Hence a proper choice for the range of $E$ ensures proper sensitivity (Smaller values of $E$ will cause very rapid fluctuations, while larger values will be monotonous.)

# 5. Neutrino Oscillation in Matter

## 5.1. Factors affecting Neutrinos in Matter

In the previous section, we calculated the oscillation probability of neutrinos travelling in vacuum. But that is usually not enough as neutrino experiments often involve detecting neutrinos that originate at a source on the earth's surface, travel through the bulk material of the earth and get detected at a detector placed on the surface several kilometres away. In such cases, we need to consider matter effects as well.

The presence of matter in a neutrinos path may affect its oscillation probability in two ways:

- Interaction with matter may cause a flavour change in a neutrino. But the Standard Model predicts that all neutrino-matter interactions are flavour conserving [3]. Hence we do not consider this possibility.

- Neutrinos can undergo forward scattering while interacting with ambient particles, which will give rise to an extra interaction potential energy. This can happen in two ways:
    1. A neutrino $\nu_\alpha$ can exchange a $W$ boson with the corresponding charged lepton $\ell_\alpha$ only. The only charged lepton that is found in the bulk earth material in significant numbers is the electron. Hence this effect will be observed in a $\nu_e$ neutrino (and its antiparticle $\bar{\nu}_e$) only. The corresponding interaction potential will obviously be proportional to the number density of electrons, $N_e$, in the bulk matter. According to the Standard Model, this potential will also be proportional to the Fermi coupling constant $G_F$ [1, 3]. The Standard Model gives,
    $$V_W = \begin{cases} +\sqrt{2} G_F N_e & \text{for } \nu_e \\ -\sqrt{2} G_F N_e & \text{for } \bar{\nu}_e \end{cases} \tag{82}$$
    2. According to the Standard Model, a neutrino of any flavour can exchange a $Z$ boson with an ambient electron, proton or neutron. Also, the $Z$-couplings to electron and proton are equal and opposite. Thus, since the bulk earth matter is electrically neutral, the electron and proton contributions via $Z$-boson exchange will cancel out. Finally, a flavour-independent potential $V_Z$ will stand out, and will be proportional to the number of neutrons per unit volume, $N_n$. The Standard Model gives [3],
    $$V_Z = \begin{cases} -\dfrac{\sqrt{2}}{2} G_F N_n & \text{for } \nu \\ +\dfrac{\sqrt{2}}{2} G_F N_n & \text{for } \bar{\nu} \end{cases} \tag{83}$$

To begin with, let's consider the Schrödinger equation in the lab-frame for a neutrino traveling through matter:
$$i \frac{\partial}{\partial t} |\nu(t)\rangle = \mathcal{H} |\nu(t)\rangle \tag{84}$$
where $|\nu(t)\rangle$ is the multi-component neutrino state vector with one component for each of the neutrino flavours [3]. For two neutrino flavours $e$ and $\mu$, for example,
$$|\nu(t)\rangle = \begin{pmatrix} f_e(t) \\ f_\mu(t) \end{pmatrix} \tag{85}$$
where $f_\alpha(t)$ is the amplitude of the neutrino being a $\nu_\alpha$ at time $t$. Therefore, $\mathcal{H}$ is a $2 \times 2$ matrix in $\nu_e$-$\nu_\mu$ space.

## 5.2. Finding the Vacuum Hamiltonian

We will first work with two-neutrino case in vacuum and find out the components of $\mathcal{H}_{\text{Vac}}$, which is the Hamiltonian corresponding to the vacuum case. To find out $(\alpha, \beta)$ component of the $\mathcal{H}_{\text{Vac}}$, we shall use (21). Starting from (21) and applying (53), the $\nu_\alpha$-$\nu_\beta$ component of $\mathcal{H}_{\text{Vac}}$ comes out to be:

$$\langle \nu_\alpha | \mathcal{H}_{\text{Vac}} | \nu_\beta \rangle = \langle \sum_i U_{\alpha i} \nu_i | \mathcal{H}_{\text{Vac}} | \sum_j U_{\beta j} \nu_j \rangle$$

$$= \sum_i \langle U_{\alpha i} \nu_i | \mathcal{H}_{\text{Vac}} | U_{\beta i} \nu_i \rangle$$

(all terms with $\langle \nu_i | \nu_j \rangle$ with $i \neq j$ vanish due to orthogonality.)

$$= \sum_i U^*_{\alpha i} U_{\beta i} E_i \langle \nu_i | \nu_i \rangle$$

$$= \sum_i U^*_{\alpha i} U_{\beta i} \sqrt{p^2 + m_i^2}$$

(86)

where we have used the fact that $E_i$ is the energy of mass eigenstate $\nu_i$.

Using (86) and the matrix in (80) we evaluate each of the four terms in $\mathcal{H}_{\text{Vac}}$. We denote the terms as $\mathcal{H}_{\alpha\alpha}$, $\mathcal{H}_{\alpha\beta}$, $\mathcal{H}_{\beta\alpha}$, and $\mathcal{H}_{\beta\beta}$ respectively. Also, we use the highly relativistic approximation $\sqrt{p^2 + m_i^2} = \left(p + \frac{m_i^2}{2p}\right)$.

$\mathcal{H}_{\alpha\alpha}$:

$$\mathcal{H}_{\alpha\alpha} = \cos^2 \theta \left(p + \frac{m_1^2}{2p}\right) + \sin^2 \theta \left(p + \frac{m_2^2}{2p}\right)$$

$$= (\cos^2 \theta + \sin^2 \theta) p + \cos^2 \theta \frac{m_1^2}{2p} + (1 - \cos^2 \theta) \frac{m_2^2}{2p}$$

$$= p + \cos^2 \theta \frac{m_1^2}{2p} + \frac{m_2^2}{2p} - \cos^2 \theta \frac{m_2^2}{2p}$$

$$= p - \frac{\cos^2 \theta}{2p} \Delta m^2 + \frac{m_2^2}{2p}$$

$$= p + \frac{m_2^2}{2p} - \frac{2 \cos^2 \theta - 1}{4p} \Delta m^2 - \frac{\Delta m^2}{4p};$$

$$= -\frac{2 \cos^2 \theta - 1}{4p} \Delta m^2 + p + \frac{2m_2^2 - \Delta m^2}{4p}$$

$$= -\cos 2\theta \frac{\Delta m^2}{4p} + p + \frac{m_1^2 + m_2^2}{4p}.$$

Proceeding similarly for $\mathcal{H}_{\beta\beta}$:

$$\mathcal{H}_{\beta\beta} = \sin^2 \theta \left(p + \frac{m_1^2}{2p}\right) + \cos^2 \theta \left(p + \frac{m_2^2}{2p}\right)$$

$$= \cos 2\theta \frac{\Delta m^2}{4p} + p + \frac{m_1^2 + m_2^2}{4p}.$$

$\mathcal{H}_{\alpha\beta}, \mathcal{H}_{\beta\alpha}$:

$$\mathcal{H}_{\alpha\beta} = \mathcal{H}_{\beta\alpha} = -\cos \theta \sin \theta \left(p + \frac{m_1^2}{2p}\right) + \sin \theta \cos \theta \left(p + \frac{m_2^2}{2p}\right)$$

$$= (-\sin \theta \cos \theta + \sin \theta \cos \theta) p - \cos \theta \sin \theta \frac{m_1^2}{2p} + \sin \theta \cos \theta \frac{m_2^2}{2p}$$

$$= \sin \theta \cos \theta \left(\frac{m_2^2}{2p} - \frac{m_1^2}{2p}\right)$$

$$= \sin 2\theta \frac{\Delta m^2}{4p}.$$

Thus,
$$\mathcal{H}_{\text{Vac}} = \frac{\Delta m^2}{4p}\begin{bmatrix} -\cos 2\theta & \sin 2\theta \\ \sin 2\theta & \cos 2\theta \end{bmatrix} + \left(p + \frac{m_1^2 + m_2^2}{4p}\right)\begin{bmatrix} 1 & 0 \\ 0 & 1 \end{bmatrix}$$
(87)

As only the relative phases of the interfering contributions matter, and consequently only the relative energies matter, we can freely subtract a multiple of the identity matrix from the expression of $\mathcal{H}_{\text{Vac}}$ [3]. This will not affect the differences between the eigenvalues of $\mathcal{H}$.

Also, for a highly relativistic neutrino, we can approximate $p \cong E$. Thus (87) becomes
$$\mathcal{H}_{\text{Vac}} = \frac{\Delta m^2}{4E}\begin{bmatrix} -\cos 2\theta & \sin 2\theta \\ \sin 2\theta & \cos 2\theta \end{bmatrix}$$
(88)

## 5.3. Finding the Hamiltonian in Matter

Now we shall construct the corresponding Hamiltonian $\mathcal{H}_M$ for propagation in matter. As discussed earlier, we expect contributions from two more factors, the interaction potentials $V_W$ and $V_Z$ [3].
$$\mathcal{H}_M = \mathcal{H}_{Vac} + V_W\begin{bmatrix} 1 & 0 \\ 0 & 0 \end{bmatrix} + V_Z\begin{bmatrix} 1 & 0 \\ 0 & 1 \end{bmatrix}.$$
(89)

- Since the $V_W$ term affects only $\nu_e$'s, only the upper left term, corresponding to $\nu_e$-$\nu_e$ element, is non-vanishing.
- Since the $V_Z$ term affects all flavours equally, a diagonal identity matrix is required.

We rewrite (89) as,
$$\mathcal{H}_M = \mathcal{H}_{\text{Vac}} + V_W\begin{bmatrix} 1 & 0 \\ 0 & -1 \end{bmatrix} + V_W\begin{bmatrix} 1 & 0 \\ 0 & 1 \end{bmatrix} + V_Z\begin{bmatrix} 1 & 0 \\ 0 & 1 \end{bmatrix}.$$

As discussed earlier, we can conveniently add or subtract any multiples of the identity matrix from the expression of the Hamiltonian without affecting the relative energy eigenvalues. Thus, the equation reduces to
$$\mathcal{H}_M = \mathcal{H}_{Vac} + V_W\begin{bmatrix} 1 & 0 \\ 0 & -1 \end{bmatrix}.$$
(90)

Using (82) and (88),
$$\mathcal{H}_M = \frac{\Delta m^2}{4E}\begin{bmatrix} -\left(\cos 2\theta - \frac{(V_W/2)}{(\Delta m^2/4E)}\right) & \sin 2\theta \\ \sin 2\theta & \left(\cos 2\theta - \frac{(V_W/2)}{(\Delta m^2/4E)}\right) \end{bmatrix}$$
$$= \frac{\Delta m^2}{4E}\begin{bmatrix} -(\cos 2\theta - x) & \sin 2\theta \\ \sin 2\theta & (\cos 2\theta - x) \end{bmatrix}.$$
(91)

where,
$$x = \frac{(V_W/2)}{(\Delta m^2/4E)} = \frac{2\sqrt{2}G_F N_e E}{\Delta m^2}$$
(92)

For simplicity, we wish to write (91) in the form of (88). It can be done if we find an $X$ such that,
$$\cos 2\theta_M = (\cos 2\theta - x)X$$
$$\sin 2\theta_M = (\sin 2\theta_M)X$$
$$\Delta m_M^2 = \frac{\Delta m^2}{X}$$

Solving these equations, we get
$$sin^2 2\theta_M = \frac{sin^2 2\theta}{sin^2 2\theta + (cos 2\theta - x)^2}$$
(93)

$$\Delta m_M^2 = \Delta m^2 \sqrt{sin^2 2\theta + (cos 2\theta - x)^2}$$
(94)

Thus,
$$\mathcal{H}_M = \frac{\Delta m_M^2}{4E}\begin{bmatrix} -\cos 2\theta_M & \sin 2\theta_M \\ \sin 2\theta_M & \cos 2\theta_M \end{bmatrix}. \quad (95)$$

So, the Hamiltonian in matter is identical to that in vacuum if $\Delta m^2$ and $\theta$ are replaced, respectively, by $\Delta m_M^2$ and $\theta_M$ using (93) and (94). Thus the effective $\Delta m^2$ and $\theta$ are different in matter from that in vacuum.

## 5.4. Finding the Oscillation Probability in Matter

From (53) and (80), we get, for the two-neutrino case (note, $\theta$ has to be replaced with $\theta_M$ in case of matter)
$$|\nu_e\rangle = |\nu_1\rangle \cos\theta_M + |\nu_2\rangle \sin\theta_M$$
$$|\nu_\mu\rangle = -|\nu_1\rangle \sin\theta_M + |\nu_2\rangle \cos\theta_M$$
$$(96)$$

The eigenvalues of $\mathcal{H}_M$ from (95) are
$$\lambda_1 = +\frac{\Delta m_M^2}{4E}, \quad \lambda_2 = -\frac{\Delta m_M^2}{4E} \quad (97)$$

Thus solving (84) with $|\nu(\tau=0)\rangle = |\nu_e\rangle$ and using (96) and (97),
$$|\nu(t)\rangle = |\nu_1\rangle e^{-i\frac{\Delta m_M^2}{4E}t} \cos\theta_M + |\nu_2\rangle e^{+i\frac{\Delta m_M^2}{4E}t} \sin\theta_M \quad (98)$$

The probability that $|\nu(t)\rangle$ is detected as $|\nu_\mu\rangle$ at time $t$ is
$$\begin{aligned}
P_M(\nu_e \to \nu_\mu) &= |\langle \nu_\mu | \nu(t)\rangle|^2 \\
&= \left| -\sin\theta_M\, e^{-i\frac{\Delta m_M^2}{4E}t} \cos\theta_M + \cos\theta_M\, e^{+i\frac{\Delta m_M^2}{4E}t} \sin\theta_M \right|^2 \\
&= \left| \sin\theta_M \cos\theta_M \left( e^{+i\frac{\Delta m_M^2}{4E}t} - e^{-i\frac{\Delta m_M^2}{4E}t} \right) \right|^2 \\
&= \left| \sin\theta_M \cos\theta_M \left( 2i \sin\frac{\Delta m_M^2}{4E}t \right) \right|^2 \\
&= \sin^2 2\theta_M \sin^2\left( \frac{\Delta m_M^2}{4E}t \right) \\
&= \sin^2 2\theta_M \sin^2\left( \Delta m_M^2 \frac{L}{4E} \right).
\end{aligned}$$
$$(99)$$

In the last step we have replaced $t$ with $L$ as we have considered the highly relativistic case.

As we can get back the case of vacuum oscillations just by replacing $\Delta m_M^2$ and $\theta_M$ with $\Delta m^2$ and $\theta$ respectively, we get from (99),
$$P(\nu_e \to \nu_\mu) = \sin^2 2\theta \sin^2\left( \Delta m^2 \frac{L}{4E} \right)$$
which is exactly (81), as expected.

## 5.5. Discussions

Clearly, matter effects are significantly determined by the quantity $x$. We would like to find out the extent to which $x$ could be significant. We shall calculate the value of $x$ for an atmospheric neutrino that travels through the earth's mantle on its way to a detector.

We know, the squared-mass splitting for atmospheric neutrinos is about $\Delta m^2 = 2.4 \times 10^{-3}$ eV² [4, 5].
The value of Fermi coupling constant $\approx 1.67 \times 10^{-23}$ eV$^{-2}$.
Density of mantle $\cong 3000$ kg/m³.

Number of (protons + neutrons) in 1 m$^3$ = $\frac{3000 \text{ kg}}{1.67 \times 10^{-27} \text{ kg}}$ = $1.80 \times 10^{30}$.

So number density of electrons, $N_e$ = number density of protons = $\frac{1.80 \times 10^{30}}{2}$ m$^{-3}$ = $8.98 \times 10^{29}$ m$^{-3}$.
(Assuming number of protons and neutrons are almost equal in mantle elements.)

From Table 1.3, we find, 1 m ≡ $5.07 \times 10^{15}$ GeV$^{-1}$.
So, 1 m$^{-3}$ = $7.67 \times 10^{-48}$ GeV$^3$.
So $N_e$ = $8.98 \times 10^{29}$ m$^{-3}$ = $7.19 \times 10^{-18}$ GeV$^3$ = $6.89 \times 10^9$ eV$^3$.

Putting this value in (92),

$$|x| = \frac{2\sqrt{2} \times 1.17 \times 10^{-23} \text{ eV}^{-2} \times 6.89 \times 10^9 \text{ eV}^3}{2.4 \times 10^{-3} \text{ eV}^2} E$$
$$= 9.50 \times 10^{-11} E \text{ eV}^{-1}$$
$$\approx \frac{E}{10.53 \text{ GeV}}.$$

(100)

The value of $x$ is proportional to the neutrino energy $E$. Thus, for a neutrino with high energy, say 20 GeV, the matter effect is large. To demonstrate the dependence, we plot $\frac{\Delta m_M^2}{\Delta m^2}$ vs. $E$ and $\frac{\sin^2 2\theta_M}{\sin^2 2\theta}$ vs. $E$ in Fig 5.1 based on equations (93) and (94).

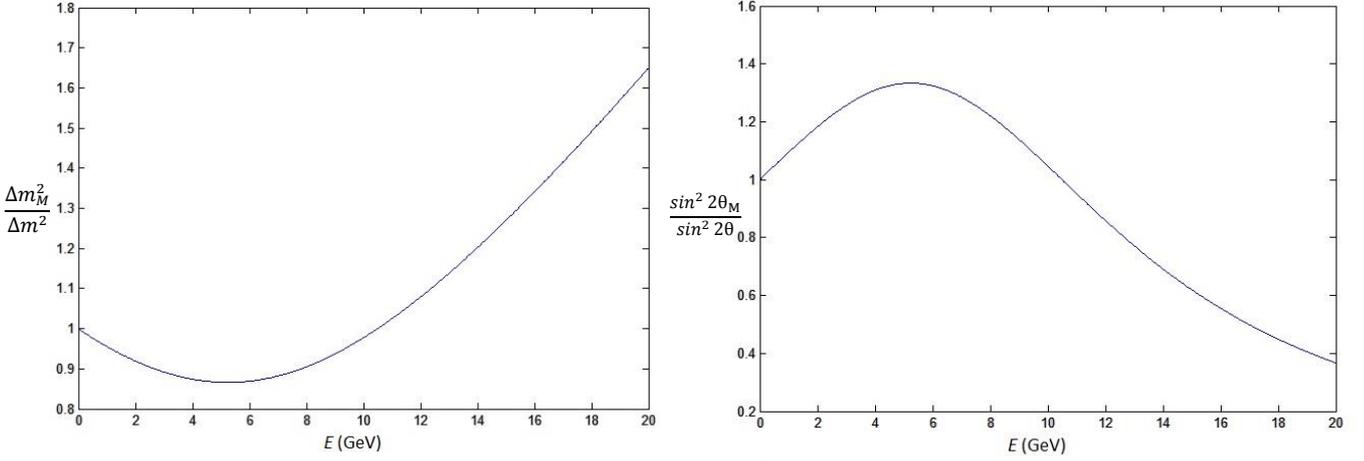

Figure 5.1: Plots for $\frac{\Delta m_M^2}{\Delta m^2}$ vs. $E$ and $\frac{\sin^2 2\theta_M}{\sin^2 2\theta}$ vs. $E$ at $\theta = 30°$.

In special cases, matter effects can be dramatically large. Consider the case where $\theta$ is very small and $x \cong \cos 2\theta$. From (93) we can see that even though $\sin^2 2\theta \ll 1$, $\sin^2 2\theta_M \cong \frac{\sin^2 2\theta}{\sin^2 2\theta} = 1$, its maximum value. Thus, the presence of matter in the path of a neutrino may cause a dramatic amplification of the mixing angle in such special cases [3]. This phenomenon, known as the Mikheyev–Smirnov–Wolfenstein (MSW) resonance effect, was believed to occur in the case of neutrinos within solar matter. However, later experiments showed that the vacuum mixing angle for solar neutrinos is larger ($\theta \cong 34°$).

As we noted earlier, $x$ will have an opposite sign if an antineutrino is used instead of a neutrino. Correspondingly, as equations (93) and (94) suggest, the values of $\Delta m_M^2$ and $\theta_M$ will be different for neutrinos and antineutrinos. This can be a method to differentiate between neutrinos and their antiparticles, but as we shall see in the next section, the relative hierarchies of the values of $m_i^2$'s and consequently the sign of the parameter $x$ is also not known.

# 6. Properties of Neutrinos

## 6.1. The Mass-squared Splittings

In the discussion till now, we had assumed in some places that there are a total of 3 mass eigenstates for notational simplicity. But we do not know the total number of mass eigenstates for sure. However, the fact that there are at least three, can be easily understood from studying solar and atmospheric neutrinos.

- Solar neutrinos have a squared-mass splitting of $\Delta m_{\text{sol}}^2 \cong 8.0 \times 10^{-5}$ eV² [6]. We attribute the corresponding splitting to $\nu_1$ and $\nu_2$. Thus, $\Delta m_{21}^2 \equiv m_2^2 - m_1^2 \cong 8.0 \times 10^{-5}$ eV².
- Atmospheric neutrinos have a squared-mass splitting of $\Delta m_{\text{atm}}^2 \cong 2.4 \times 10^{-3}$ eV² [4, 5]. We must have one more mass eigenstate $\nu_3$ to accommodate for this. So we attribute the atmospheric splitting to $\nu_2$ and $\nu_3$. Thus, $\Delta m_{32}^2 \equiv m_3^2 - m_2^2 \cong 2.4 \times 10^{-3}$ eV².

In the above representation, we have assumed $m_3^2 \gg m_2^2 > m_1^2$. However, as (91) indicates, the value of $x$, the parameter that determines matter effects in neutrino oscillations, reverses sign for a particular mass pair if the corresponding hierarchy is reversed. That is, if we do not assume a definite hierarchy, we cannot tell apart a neutrino and an antineutrino, and vice versa.

As discussed earlier, the $\Delta m^2$ sensitivity of an experiment can be regulated by regulating the value of $\frac{L}{E}$. Thus to detect solar neutrinos, we might have a setup with $E \approx 1$ GeV and $L \approx 10^5$ km. Whereas, for atmospheric neutrinos, a setup with $E \approx 1$ GeV and $L \approx 10^3$ km would be ideal.

What if there are more than three mass eigenstates? If such mass eigenstates exist, their linear combinations will give rise to one or more flavours of neutrinos each of which does not form an isospin doublet with a lepton. Thus they cannot couple to $W$ or $Z$ bosons unlike the three observed neutrino flavours. This implies, they cannot participate in weak interactions. Such neutrinos are called sterile neutrinos [7].

## 6.2. The Flavour and Mass States and Mixing

We know, each of the flavour states is a superposition of mass eigenstates, where the fraction of $\nu_i$ in $\nu_\alpha$ is given by $|U_{\alpha i}|^2$, or equivalently, fraction of $\nu_\alpha$ in $\nu_i$ is $|U_{\alpha i}|^2$.

Experimentally, it turns out that $|U_{e3}|^2$ is very small [8]. A quantity $\theta_{13}$ defined as $|U_{e3}|^2 = \sin^2 \theta_{13}$ is the corresponding mixing angle. Experiments show that this mixing angle has a value of ~9°, indicating very little mixing [9]. Thus, $\nu_3$ is essentially superposition of only $\nu_\mu$ and $\nu_\tau$. This $\nu_\mu$-$\nu_\tau$ mixing angle is, however, large (~45°), indicating maximal mixing.

For $\nu_2$, $|U_{e2}|^2 \approx |U_{\mu 2}|^2 \approx |U_{\tau 2}|^2 \approx \frac{1}{3}$. For $\nu_1$, $|U_{e1}|^2 \approx \frac{2}{3}$, $|U_{\mu 2}|^2 \approx |U_{\tau 2}|^2 \approx \frac{1}{6}$. [3]

Table 6.1 summarizes the above experimental observations.

*Table 6.1: Table indicating mixing probabilities in neutrinos*

|  | $\nu_e$ | $\nu_\mu$ | $\nu_\tau$ |
|---|---|---|---|
| $\nu_3$ | $|U_{e3}|^2 \ll 1$ | $|U_{\mu 3}|^2 \approx \frac{1}{2}$ | $|U_{\tau 3}|^2 \approx \frac{1}{2}$ |
| $\nu_2$ | $|U_{e2}|^2 \approx \frac{1}{3}$ | $|U_{\mu 2}|^2 \approx \frac{1}{3}$ | $|U_{\tau 2}|^2 \approx \frac{1}{3}$ |
| $\nu_1$ | $|U_{e1}|^2 \approx \frac{2}{3}$ | $|U_{\mu 1}|^2 \approx \frac{1}{6}$ | $|U_{\tau 1}|^2 \approx \frac{1}{6}$ |

Recalling that the matrix $U$ is a $3 \times 3$ unitary matrix, we should be able to rewrite $U$ in terms of (at least) 3 rotation angle and 1 phase factor. It is useful to represent $U$ as [3],

$$U = \underbrace{\begin{bmatrix} 1 & 0 & 0 \\ 0 & c_{23} & s_{23} \\ 0 & -s_{23} & c_{23} \end{bmatrix}}_{\text{Atmospheric}} \times \underbrace{\begin{bmatrix} c_{13} & 0 & s_{13}e^{i\delta} \\ 0 & 1 & 0 \\ -s_{13}e^{i\delta} & 0 & c_{33} \end{bmatrix}}_{\text{Cross}} \times \underbrace{\begin{bmatrix} c_{12} & s_{12} & 0 \\ -s_{12} & c_{12} & 0 \\ 0 & 0 & 1 \end{bmatrix}}_{\text{Solar}}.$$

Note, the four independent factors are $\theta_{12}, \theta_{23}, \theta_{13}$ and $\delta$, and $c_{12}$ represents $\cos\theta_{12}$, and $s_{12} \equiv \sin\theta_{12}$.

Approximating atmospheric oscillation as a two neutrino $\nu_\mu$-$\nu_\tau$ problem, $\theta_{23} \approx \theta_{\text{atm}}$. This is a valid approximation as $\nu_e$ is almost inexistent in $\nu_3$, which participates in atmospheric oscillation. Turns out, this $\theta_{\text{atm}}$ is approximately $45°$, implying maximal mixing [4, 5].

Similarly we can approximate oscillation as a two neutrino problem, $\theta_{12} \approx \theta_{\text{sol}}$. This is a valid approximation as $\nu_e$ is essentially either $\nu_1$ or $\nu_2$ and solar neutrinos are born as $\nu_e$ only. Experimentally, $\theta_{\text{sol}} \approx 33.9°$ indicating large, though not maximal, mixing [6].

The cross-mixing matrix contains a CP violating term $\delta$. However, since it is multiplied with $s_{13}e^{i\delta}$, its effect is negligible as long as $\theta_{13}$ is zero or very small. As discussed earlier, $\theta_{13} \approx 9°$ [8, 9], which is small, but cannot be ignored.

# 7. Conclusions

As we have seen, assuming the existence of non-degenerate mass eigenstates of neutrinos gives us a probability-based model that generously accommodates for the experimentally observed phenomenon of neutrino oscillation. The existence of the phenomenon itself refutes the assumption that neutrinos are massless, and makes us look beyond the Standard Model. Numerous experiments in the recent decades have not only helped us reconfirm the phenomenon, but also have helped us determine the related parameters with better and better accuracies.

However, neutrino physics is still an active area of research with many more open questions and unsolved problems. We are in the process of building yet more sensitive detectors to make more precise measurements and observations.

One such open question, for example, is whether more than three mass eigenstates exist. As discussed, that would imply the presence of sterile neutrinos. The LSND experiment data suggests presence of another mass-squared splitting in the range of $0.2\sim10$ eV$^2$ [7]. To acccommodate for this splitting we shall definitely require one more mass state that is separated by a large magnitude from the other three.

We would also like to know the explicit values of the masses of the mass eigenstates. As discussed earlier, neutrino oscillation experiments can only give the relative squared-splittings of these values, from which we can, at best, find a range of values for the masses of the mass eigenstates.

Is there any difference between a neutrino and its antineutrino? We would like to know whether neutrinos are Majorana particles (particle identical to its antiparticle) or Dirac particles (particles and antiparticles are distinct). For this we might want to find out whether a neutrinoless double beta decay is possible. A $\beta$-decay refers to the process $n \to p + e^- + \bar{\nu}_e$. A neutrinoless double beta decay would refer to the process, $2n \to 2p + 2e^-$. To explain this, we must introduce two virtual $\bar{\nu}_e$'s for each of the beta decay processes. But since the $\bar{\nu}_e$ term doesn't appear in the final product, each of them must have annihilated the other, which would imply that at least one flavour of the neutrino is a Majorana particle.

Thus the phenomenon of neutrino oscillation and the discovery of neutrino mass has brought along with them several unsolved problems in particle physics, and we would obviously want to find ways to solve these.